\documentclass[apj]{emulateapj}

\def\ms{\,m\,s$^{-1}$}         
\def\kms{\,km\,s$^{-1}$}       
\def\msol{$M_\odot$}		
\def\rsol{$R_\odot$}		
\def\denssol{$\rho_\odot$}	
\def\mstar{$M_*$}		
\def\rstar{$R_*$}		
\def\densstar{$\rho_*$}		
\def\mplanet{$M_{\rm P}$}	
\def\rplanet{$R_{\rm P}$}	
\def\densplanet{$\rho_{\rm P}$}
\def\mjup{$M_{\rm Jup}$}	
\def\densjup{$\rho_{\rm Jup}$}	
\def\teql{$T_{\rm eql}$}
\def\teff{$T_{\rm eff}$}
\def\feh{[Fe/H]}
\def\logg{$\log g_*$}
\def\vsini{$v_* \sin I_*$}
\def\mictrb{$\xi_{\rm t}$}

\def\kms{km\, s$^{-1}$}

\def\svsicos{$\sqrt{v_* \sin I} \cos \lambda$}
\def\svsisin{$\sqrt{v_* \sin I} \sin \lambda$}

\def\rhk{$\log R'_{\rm HK}$}

\shorttitle{The well-aligned orbit of WASP-84b}
\shortauthors{D. R. Anderson et al.}

\begin{document}


\title{The well-aligned orbit of WASP-84\lowercase{b}: evidence for disc migration\altaffilmark{$\dagger$}}


\author{D.~R.~Anderson\altaffilmark{1},
A.~H.~M.~J.~Triaud\altaffilmark{2},
O.~D.~Turner\altaffilmark{1},
D.~J.~A.~Brown\altaffilmark{3,4},
B.~J.~M.~Clark\altaffilmark{1},
B.~Smalley\altaffilmark{1},
A.~Collier~Cameron\altaffilmark{5},
A.~P.~Doyle\altaffilmark{1},
M.~Gillon\altaffilmark{6},
C.~Hellier\altaffilmark{1},
C.~Lovis\altaffilmark{7},
P.~F.~L.~Maxted\altaffilmark{1},
D.~Pollacco\altaffilmark{3},
D.~Queloz\altaffilmark{7,8},
A.~M.~S.~Smith\altaffilmark{9,1},
}

\email{d.r.anderson@keele.ac.uk}

\affil{$^1$ Astrophysics Group, Keele University, Staffordshire ST5 5BG, UK\\
$^2$ Department of Physics, and Kavli Institute for Astrophysics and Space Research, MIT, Cambridge, MA 02139, USA\\
$^3$ Department of Physics, University of Warwick, Coventry CV4 7AL, UK\\
$^4$ Astrophysics Research Centre, School of Mathematics \& Physics, Queen's University, University Road, Belfast BT7 1NN\\
$^5$ SUPA, School of Physics and Astronomy, University of St. Andrews, North Haugh, Fife KY16 9SS, UK\\
$^6$ Institut d'Astrophysique et de G\'eophysique,  Universit\'e de Li\`ege,  All\'ee du 6 Ao\^ut, 17,  Bat.  B5C, Li\`ege 1, Belgium\\
$^7$ Observatoire de Gen\`eve, Universit\'e de Gen\`eve, 51 Chemin des Maillettes, 1290 Sauverny, Switzerland\\
$^8$ Cavendish Laboratory, J J Thomson Avenue, Cambridge, CB3 0HE, UK\\
$^9$ N. Copernicus Astronomical Centre, Polish Academy of Sciences, Bartycka 18, 00-716, Warsaw, Poland\\
}

\altaffiltext{$^\dagger$}{Based on observations made with: 
  the HARPS-North spectrograph on the 3.6-m Telescopio Nazionale Galileo under OPTICON program 2013B/069; 
  the HARPS spectrograph on the ESO 3.6-m telescope under program 090.C-0540; 
  and the RISE photometer on the 2.0-m Liverpool Telescope under programs PL12B13 and PL14A11. 
  The photometric time-series and radial-velocity data used in this work are available at the CDS.}



\begin{abstract}
We report the sky-projected orbital obliquity (spin-orbit angle) of WASP-84b, a 0.70-\mjup\ planet in a 8.52-day orbit around a G9V/K0V star, to be $\lambda = 0.3 \pm 1.7^\circ$. 
We obtain a true obliquity of $\psi = 14.8 \pm 8.0^\circ$
from a measurement of the inclination of the stellar spin axis with respect to the sky plane. 
Due to the young age and the weak tidal forcing of the system, we suggest that the orbit of WASP-84b is unlikely to have both realigned and circularised from the misaligned and/or eccentric orbit likely to have arisen from high-eccentricity migration. Therefore we conclude that the planet probably migrated via interaction with the protoplanetary disc. This would make it the first short-orbit, giant planet to have been shown to have migrated via this pathway. 
Further, we argue that the distribution of obliquities for planets orbiting cool stars (\teff\ $<$ 6250 K) suggests that high-eccentricity migration is an important pathway for the formation of short-orbit, giant planets.
\end{abstract}


\keywords{planets and satellites: dynamical evolution and stability --- 
planet-disk interactions --- 
planet-star interactions --- 
planets and satellites: individual: WASP-84b --- 
stars: individual: WASP-84}



\section{Introduction}
The orbital obliquity (spin-orbit angle; $\psi$) distribution of short-orbit, giant planets, or  ``hot Jupiters'', may be indicative 
of the manner in which they arrived in their current orbits from farther out, where 
they presumably formed (e.g. \citealt{2006ApJ...648..666R}).
As a star and its planet-forming disc both inherit their angular momenta from their parent molecular cloud, 
stellar spin and planetary orbital axes are expected to be, at least initially, aligned ($\psi = 0$). 
Migration via interaction with the gas disc is expected to preserve this initial spin-orbit alignment \citep{1996Natur.380..606L, 2009ApJ...705.1575M}.
Migration via high-eccentricity migration, in which a cold Jupiter is perturbed into an eccentric, misaligned orbit that is then circularised, shortened and realigned by tidal dissipation, is expected to produce a broad range of obliquities \citep{2007ApJ...669.1298F, 2008ApJ...678..498N, 2010ApJ...725.1995M, 2011Natur.473..187N}.

A broad range of obliquities has been found for hot-star systems (\teff\ $>$ 6250\,K), for which 
tidal realignment is expected to be inefficient due to the absence of a 
substantial convective envelope \citep{2010ApJ...718L.145W,2010ApJ...719..602S}, whereas systems with massive stars that had no convective envelope on the main sequence, but that are old enough to have developed convective envelopes ($\gtrsim$2.5\,Gyr), are aligned \citep{2011A&A...534L...6T}.
Conversely, cool-star systems experiencing strong tidal forcing---that is those with short scaled semi-major axes, 
$a$/\rstar, and high planet-to-star mass ratios---tend to be aligned (see \citealt{2012ApJ...757...18A} and references therein). 
The obliquities of nine planets orbiting cool stars and experiencing weak tidal forcing ($a$/\rstar $>$ 15) have been 
measured: 
HAT-P-11b, 
\citep{2010ApJ...723L.223W,2011PASJ...63S.531H,2011ApJ...743...61S}, 
HAT-P-17b 
\citep{2013ApJ...772...80F}, 
HAT-P-18b 
\citep{2014A&A...564L..13E}, 
HD\,17156\,b 
\citep{2008PASJ...60L...1N, 2008ApJ...683L..59C, 2009A&A...503..601B, 2009PASJ...61..991N}, 
HD\,80606\,b 
\citep{2009A&A...498L...5M, 2009A&A...502..695P, 2009ApJ...703.2091W, 2010A&A...516A..95H}, 
Kepler-30b 
\citep{2012Natur.487..449S}, 
Kepler-63b 
\citep{2013ApJ...775...54S}, 
WASP-8b 
\citep{2010A&A...517L...1Q} and 
WASP-117b \citep{2014A&A...568A..81L}. 
The orbits of all but HD\,17156\,b and, possibly, HAT-P-17b are misaligned and all but HAT-P-18b and Kepler-63b are eccentric. 

Various groups are attempting to reproduce the observed obliquity distribution with models (e.g. \citealt{2012ApJ...754L..36N, 2013ApJ...769L..10R, 2014ApJ...784...66X, 2014ApJ...790L..31D}).
We require a larger sample of measured obliquities, especially for weak-tide, cool-star systems, and more realistic models to be able to discern the relative contribution of the different migration pathways and to better understand the physical processes involved in tidal dissipation and orbital realignment. 
Here we present an obliquity measurement for the weak-tide, cool-star system WASP-84 from observations of its  Rossiter-McLaughlin (RM) effect (e.g. \citealt{2012ApJ...757...18A}). \citet[][hereafter A13]{2013arXiv1310.5654A} found the WASP-84 system to comprise a 0.69-\mjup\ planet in a circular ($e < 0.077$ at 2\,$\sigma$), 8.52-d orbit around an active K0V star. 

\section{Observations}
We obtained 17 spectra of WASP-84 with HARPS on the ESO 3.6-m telescope \citep{2002Msngr.110....9P} through the transit of 2013 Feb 4--5 
and a further 13 spectra around the orbit.
The transit spectra, with exposure times of 1200\,s, were taken over an airmass range of 1.39--1.17--1.92 and have signal-to-noise ratio (SNR) of 19--40 per pixel at 5500\AA. 
The seeing deteriorated to $>$2\arcsec\ going into the transit, which reduced the flux entering the fibre. 
We chose to switch the read-out mode from fast to slow for the final 6 spectra, aiming for a higher precision at the expense of time resolution.
The Moon, 31 per cent illuminated and 115$^\circ$ from WASP-84, rose at the time of mid-transit. 
The spectra around the orbit had exposure times of 600\,s, except for the final spectrum, for which the exposure time was 900\,s. 

We obtained 38 spectra of WASP-84 with HARPS-North on the 3.6-m Telescopio Nazionale Galileo \citep{2012SPIE.8446E..1VC} through the transit of 2014 Jan 11--12. 
There was light cloud throughout the night, though it cleared during the latter portion of the sequence. At the time of mid-transit the Moon was 85 per cent illuminated at a distance of 66$^\circ$ from WASP-84.
We discarded the final 5 spectra as they were taken when the target was significantly beyond an airmass of 2 and the final spectrum was aborted. 
The remaining 33 spectra, each with an exposure time of 600\,s, covered an airmass range of 1.27--1.12--2.06 and have SNR of 25--42 per pixel at 5500\AA. 

We used the HARPS pipeline to process the HARPS and HARPS-North spectra and to compute radial velocities (RVs) by weighted cross-correlation with a numerical K5-spectral template \citep{2005Msngr.120...22P}. 

Concurrently with HARPS-North, we observed WASP-84 during the same transit and from the same site with the RISE fast-readout photometer mounted on the 2-m Liverpool Telescope \citep{2008SPIE.7014E.217S}. 
RISE has a single fixed $V$+$R$ filter, a field of view of 9\farcm2 $\times$ 9\farcm2 and zero read-out overhead.
We defocussed the instrument and used autoguiding to minimise the effect of flat-fielding errors.
We used an exposure time of 4\,s to acquire 4236 images. 
Over the sequence the airmass of the target ranged over 1.12--1.80. 
We performed differential aperture photometry on the images using 5 comparison stars. 

\begin{figure}
\includegraphics[width=90mm]{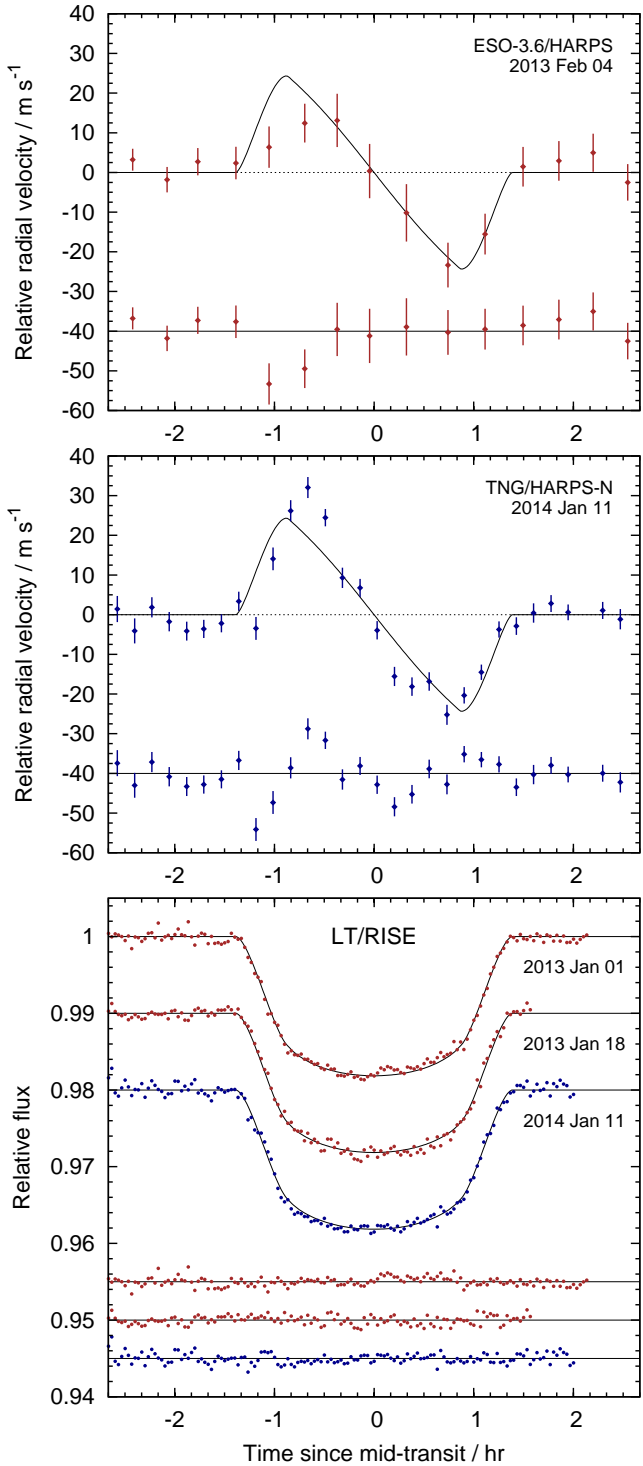}
\caption{
{\it Top panel:} 
The HARPS radial velocities with the best-fitting Rossiter-McLaughlin effect. 
The circular Keplerian model has been subtracted. 
The seeeing deteriorated going into the transit. 
{\it Middle panel:} 
The HARPS-North radial velocities with the best-fitting Rossiter-McLaughlin effect. 
The circular Keplerian model has been subtracted. 
The time-correlated noise correlates with cloud. 
{\it Bottom panel:} LT/RISE transit lightcurves presented in A13 (top two lightcurves) and herein (bottom lightcurve), offset for clarity and binned with a bin width of two minutes. 
The best-fitting transit model is superimposed. 
The residuals about the model are plotted below the lightcurves in the same order.
The planet appears to have crossed an active region shortly after mid-transit during the transit of 2013 Jan 01, which was a photometric night. The same may have happened during the ingress of the transit of 2014 Jan 11, though there was light cloud that night.
\label{fig:w84-rise-rm}} 
\end{figure}

\begin{figure}
\includegraphics[width=90mm]{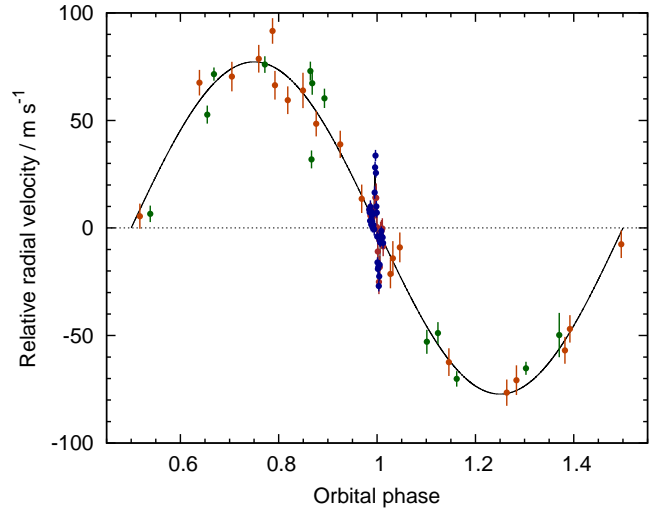}
\caption{
The radial velocities from CORALIE (pre-whitened; orange symbols; A13), HARPS (green symbols for the orbit, brown for the transit; this paper) and HARPS-North (blue symbols; this paper). 
The best-fitting circular Keplerian and RM model is superimposed. 
\label{fig:w84-rv-orb}} 
\end{figure}

We plot the HARPS and HARPS-North RVs and the RISE photometry obtained through the transit in Figure~\ref{fig:w84-rise-rm} and all the RVs around the orbit in Figure~\ref{fig:w84-rv-orb}. 
Time-correlated noise is evident in the early portion of the transit in both the HARPS and the HARPS-North RV sequences. We suggest that this results from a meteorological coincidence rather than being due to an astrophysical source: the seeing deteriorated at the start of the transit observed by HARPS, which resulted in a drop in SNR; and the scatter in the HARPS-North RVs correlates well with the cloud visible in the Liverpool Telescope's sky cameras\footnote{\url{http://telescope.livjm.ac.uk/Reports} - see especially the ``SkyCam-T'' video.}. 

\section{Stellar parameters from the HARPS spectra}
\label{sec:stellar}
We coadded the individual HARPS spectra to produce a single spectrum with an average signal-to-noise ratio of 100:1.
We performed the spectral analysis using the methods detailed in \citet{2013MNRAS.428.3164D}. 
The excitation balance of the Fe~{\sc i} lines was used to determine the effective temperature (\teff). The surface gravity (\logg) was determined from the ionisation balance of Fe~{\sc i} and Fe~{\sc ii} and using the Na~{\sc i} D lines. The iron abundance was obtained from equivalent width measurements of 47 unblended  Fe~{\sc i} and Fe~{\sc ii} lines. The quoted error estimates include that given by the uncertainties in \teff\ and \logg, as well as the scatter due to measurement and atomic data uncertainties.
The projected stellar rotation velocity (\vsini) was determined by fitting the profiles of several unblended Fe~{\sc i} lines in the wavelength range 6000--6200\AA. A value for macroturbulent velocity of 2.13 $\pm$ 0.73 \kms\ was assumed from the asteorseismic-based calibration of \citet{2014arXiv1408.3988D}.
Using those spectra with SNR $>$ 20 and assuming $B - V = 0.82$, we determined the \rhk\ activity index from the emission in the cores of the Ca~{\sc ii} H+K lines (e.g. \citealt{2009A&A...495..959B}). 
The results of the spectral analysis are given in Table~\ref{tab:params}. 

\section{Obliquity and system parameters from the HARPS radial velocities and the RISE lightcurves}
\label{sec:params}
We determined the sky-projected obliquity and the system parameters from a simultaneous fit to the transit lightcurves and the radial velocities. 
The fit was performed using the current version of the Markov-chain Monte Carlo (MCMC) code described by \citet{2007MNRAS.380.1230C} and \citet{2014arXiv1402.1482A}.
The RM effect was modelled using the formulation of \citet{2011ApJ...742...69H}.

The available time-series photometry on WASP-84 are: the WASP lightcurves spanning 2009 Jan--2011 Apr (A13); a partial-transit lightcurve from TRAPPIST (A13); two full-transit lightcurves from RISE (A13); and a third RISE lightcurve (this paper). 
The available RV data are: 20 CORALIE RVs around the orbit, pre-whitened for stellar activity using a simple harmonic series (A13); and 13 HARPS spectra around the orbit, 17 HARPS spectra through a transit, and 38 HARPS-North spectra through a transit (this paper). 
Unlike with the CORALIE RVs, we found no reason to pre-whiten the HARPS orbital RVs: the Pearson correlation coefficient between residual RV and bisector span is $r = -0.23$ for the 13 RVs; this becomes $r = 0.04$ when excluding the latest RV, which was taken the season after the other 12 measurements. This compares with $r = -0.71$ for the 20 CORALIE RVs (A13). 

In our final MCMC analysis we opted to determine the system parameters using only the three high-quality RISE lightcurves and all the RVs, which were partitioned as listed above to allow for instrumental and astrophysical offsets. 
The excluded WASP photometry was imperfectly detrended for rotational modulation and instrumental noise and the excluded TRAPPIST photometry covers only the first half of the transit and was obtained at high airmass (1.47 at ingress to 2.40 at the end of the observation).
We included the timing information of the WASP and TRAPPIST photometry in our final analysis by placing a normal prior on the time of mid-transit ($T_{\rm 0} = 2\,456\,439.52878 \pm 0.00015$ d), which was the $T_{\rm 0}$ from an initial anlysis including all the listed data, but with an error bar larger by a factor of 2 to account for the double-weighting of the remaining photometry. 

The stellar density reported in A13 (\densstar\ = $2.02 \pm 0.07$\,\denssol) seemed high a star of the reported mass ($0.84 \pm 0.04$\,\msol), which was determined from an empirical mass calibration. As a consistency check, we opted to determine stellar mass from a comparison with stellar models.
We determined \densstar\ = 1.851 $\pm$ 0.049 \denssol\ from an initial analysis that omitted the WASP and TRAPPIST data. We input that value of \densstar\ and the values of \teff\ and \feh\ from the spectral analysis into the {\sc bagemass} stellar evolution MCMC code of \citet{bagemass}, from which we obtained a stellar mass of \mstar\ = 0.855 $\pm$ 0.028 \msol\ and a stellar age of 2.0 $\pm$ 1.6 Gyr.
In our final analysis, we drew a value of \mstar\ at each MCMC step from a normal distribution with mean and standard deviation equal to the {\sc bagemass}-derived values, but with an error bar larger by a factor 2 to allow for uncertainties due to the unknown helium abundance and the effects of magnetic activity on the mass-radius relation; 
thus we had no need of a mass calibration.

In our final analysis, we adopted a circular orbit, which \citet{2012MNRAS.422.1988A} advocate is the prudent choice for hot Jupiters in the absence of evidence to the contrary. In an initial analysis, in which we fit for eccentricity, $e$, we found $e = 0.018 \pm 0.011$ and $e < 0.041$ at the 2-$\sigma$ level. 

The median values and the 1-$\sigma$ limits of our MCMC parameters' posterior distributions are given in Table~\ref{tab:params}. 
The best fits to the radial velocities and the photometry are plotted in Figures~\ref{fig:w84-rise-rm} and \ref{fig:w84-rv-orb}.
The presented solution supersedes that of the discovery paper owing to the additional RISE lightcurve and HARPS RVs, and the omission of the lower-quality WASP and TRAPPIST lightcurves.

We obtained $\lambda = 0.3 \pm 1.7^\circ$ and \vsini\ = 2.593 $\pm$ 0.087 \kms\ when fitting the RM effect to both the HARPS-North and HARPS RVs. 
Unsurprisingly, the HARPS-North RVs, with both a cadence and a typical precision twice that of the HARPS RVs, do most to constrain the RM effect: we obtained $\lambda = -0.9 \pm 1.6^\circ$ and \vsini\ = 2.650 $\pm$ 0.089 \kms\ when using only the HARPS-North RVs, and we obtained $\lambda = 11.6 \pm 7.5^\circ$ and \vsini\ = 1.96 $\pm$ 0.31 \kms\ when using only the HARPS RVs. 
Primarily due to the omission of the WASP photometry, we obtained a lower stellar density as compared to the discovery paper (1.89 $\pm$ 0.05 \denssol\ cf. 2.02 $\pm$ 0.07 \denssol; A13).

\begin{table} 
\caption{System parameters from the spectral and the MCMC analyses} 
\scriptsize
\label{tab:params}
\begin{tabular}{lc}
\hline
\hline
Parameter, Symbol / Unit & Value \\
\hline
Spectral analysis: \\
\hline
Stellar effective temperature, \teff\ / K & 5280 $\pm$ 80 \\
Stellar surface gravity, \logg\ / (cgs) & 4.65 $\pm$ 0.17 \\
Stellar metallicity, \feh\ & +0.09 $\pm$ 0.12 \\
Microturbulence, \mictrb\ / \kms & 0.7 $\pm$ 0.3 \\
Macroturbulence$^a$, $v_{\rm mac}$ / \kms & 2.13 $\pm$ 0.73 \\
Proj. stellar rot. vel., \vsini\ / \kms & 2.9 $\pm$ 0.8 \\
Lithium abundance, $\log A$(Li) & $<$0.7 \\
Ca {\sc ii} H+K activity index, \rhk & $-4.44$ \\
\hline
MCMC proposal parameters: \\
\hline
Period, $P$ / d		& 8.5234964 $\pm$ 0.0000036 \\
Mid-transit$^b$, $T_{\rm 0}$ / d		& 6448.052287 $\pm$ 0.000067 \\
Transit duration, $T_{\rm 14}$ / d	& 0.11536 $\pm$ 0.00039 \\
Transit depth, $\Delta F=R_{\rm P}^{2}$/R$_{*}^{2}$ & 0.01705 $\pm$ 0.00011 \\
Impact parameter, $b$ 				& 0.6534 $\pm$ 0.0084 \\
Reflex velocity, $K_{\rm 1}$ / \ms	& 77.2 $\pm$ 2.1 \\
Systemic velocity, $\gamma$ / \ms		& $-11\,578.2 \pm 1.5$ \\
HARPS-North offset, $\gamma_{\rm off,N,tran}$ / \ms & $-21.50 \pm 0.50$ \\
HARPS transit offset, $\gamma_{\rm off,S,tran}$ / \ms & $28.24 \pm 0.71$ \\
HARPS orbital offset, $\gamma_{\rm off,S,orb}$ / \ms & $14.00 \pm 0.44$ \\
\svsicos			& 1.610 $\pm$ 0.027 \\
\svsisin			& 0.001 $\pm$ 0.047 \\
\medskip
eccentricity, $e$					& 0 (adopted; $<$0.041 at 2\,$\sigma$) \\
\hline
MCMC derived parameters: \\
\hline
Projected orbital obliquity, $\lambda$ / $^\circ$	& $0.3 \pm 1.7$ \\
Orbital obliquity$^c$, $\psi$ / $^\circ$ & $14.8 \pm 8.0$ \\
Proj. stellar rot. vel., \vsini\ / \kms		& 2.593 $\pm$ 0.087 \\
Stellar rotation velocity$^b$, $v_*$ / \kms & 2.70 $\pm$ 0.09 \\
Stellar spin inclination$^b$, $I_*$ / $^\circ$ & 73.5 ($>$ 70.0 at 1\,$\sigma$) \\
Orbital inclination, $i_{\rm P}$	/ $^\circ$	& 88.275 $\pm$ 0.037 \\
Scaled orbital separation, $a$/\rstar 			& 21.70 $\pm$ 0.19 \\
Ingress/egress duration, $T_{\rm 12}(=T_{\rm 34})$ / d	& 0.02181 $\pm$ 0.00048 \\
Stellar mass, \mstar\ / \msol		& 0.854 $\pm$ 0.057 \\
Stellar radius, \rstar\ / \rsol		& 0.768 $\pm$ 0.018 \\
Stellar surface gravity, $\log g_{*}$ / (cgs)	& 4.598 $\pm$ 0.012 \\
Stellar density, \densstar			& 1.886 $\pm$ 0.050 \\
Planetary mass, \mplanet 			& 0.700 $\pm$ 0.037 \\
Planetary radius, \rplanet 			& 0.975 $\pm$ 0.025 \\
Planetary surface gravity, $\log g_{\rm P}$ / (cgs)	& 3.226 $\pm$ 0.016 \\
Planetary density, \densplanet\ / \densjup	& 0.755 $\pm$ 0.038 \\
Orbital major semi-axis, $a$ / au			& 0.0775 $\pm$ 0.0017 \\
Planetary equilib. temperat., \teql\ / K			& 832 $\pm$ 13 \\
\\ 
\hline 
\end{tabular} 
$^a$ $v_{\rm mac}$ value obtained using the calibration of \citet{2014arXiv1408.3988D}.\\
$^b$ $T_{\rm 0}$ is in HJD (UTC) and 2\,450\,000 has been subtracted.\\
$^c$ See Section~\ref{sec:res-disc} for the calculation of $\psi$, $v_*$ and $I_*$.
\end{table}

\section{Results and Discussion}
\label{sec:res-disc}
We find the sky-projected spin-orbit angle, or projected obliquity, of WASP-84b to be $\lambda = 0.3 \pm 1.7^\circ$. 
With a measurement of the angle between the stellar spin axis and the line of sight, $I_*$, we can calculate the true obliquity, $\psi$, using equation 9 of \citet{2009ApJ...696.1230F}). 
Using the stellar rotation period of $P_{\rm rot} = 14.36 \pm 0.35$ d, derived by A13 from observed photometric modulation, and our MCMC posterior distributions of \rstar, \vsini\, $i_{\rm P}$ and $\lambda$, we calculate $v_*$ = 2.70 $\pm$ 0.09 \kms, $I_* = 73.5^\circ$ ($I_* > 70.0^\circ$ at the 1-$\sigma$ level), and $\psi = 14.8 \pm 8.0^\circ$; these are our adopted values. 
This is consistent with a well-aligned orbit and certainly excludes polar or retrograde orbits as observed for some other hot-Jupiter systems.
We obtained consistent results from analyses using: 
only the HARPS-North RVs ($I_* = 78.5^\circ$, $I_* > 73.4^\circ$ at 1\,$\sigma$; $\psi = 10.0 \pm 7.6^\circ$); 
only the HARPS RVs ($I_* = 46.8^\circ$, $I_* > 42.2^\circ$ at 1\,$\sigma$; $\psi = 43 \pm 10^\circ$); 
all the RVs and the spectral \vsini\ ($I_* = 90.0^\circ$, $I_* > 68.7^\circ$ at 1\,$\sigma$; $\psi = 3 \pm 18^\circ$).
Using the relation between $P_{\rm rot}$ and \rhk\ of \citet{2008ApJ...687.1264M}, we obtained $P_{\rm rot} = 11.4 \pm 1.7$\,d, which is consistent with the $P_{\rm rot}$ from photometric modulation. 

We found the orbital eccentricity of WASP-84b to be low and consistent with a circular orbit: $e = 0.018 \pm 0.011$ and $e < 0.041$ at the 2-$\sigma$ level.
Using the \rhk\ age-activity relation of \citet{2008ApJ...687.1264M} we obtained an age of $0.70 \pm 0.18$\,Gyr. 
This compares with our evolutionary-analysis age of 2.1 $\pm$ 1.6\,Gyr (Section~\ref{sec:params}) and the gyrochronolgical ages of 0.8 $\pm$ 0.1\,Gyr and $\sim$1.4\,Gyr from A13. 
Using equation 2 of \citet{2012ApJ...757...18A} we calculate the relative timescale for alignment via tidal dissipation to be $4.2 \times 10^{14}$ yr. 

Considering the young age and the weak tidal forcing of the WASP-84 system, we suggest that it is improbable that the orbit of WASP-84b could have circularised and re-aligned from the eccentric, misaligned orbit likely to have arisen from migration via a high-eccentricity pathway. 
We suggest that WASP-84b probably migrated to its current 8.52-d orbit via interaction with the protoplanetary disc. This would make it the first short-orbit, giant planet to have been shown to have done so. 

Another candidate for disc migration is HAT-P-17b, which has a slightly larger scaled semi-major axis (a/\rstar\ = 22.6 cf. 21.6) and a slightly smaller mass ratio (\mplanet/\mstar\ = 0.00059 cf. 0.00078).
Though HAT-P-17b may be aligned ($\lambda = 19 \pm 15^\circ$), \citet{2013ApJ...772...80F} find that the data weakly favour a misaligned orbit. Further, the orbit is eccentric ($e = 0.342 \pm 0.006$) and the system is old (7.8 $\pm$ 3.3 Gyr; \citealt{2012ApJ...749..134H}). Thus the evidence for disc migration is less compelling.

\section{The dependence of alignment on orbital distance}
A broad range of obliquities has been found for hot-star systems (\teff\ $>$ 6250\,K), whereas cool-star systems tend to be aligned. 
This has been interpreted as evidence for realignment by tidal dissipation, which is suggested to be efficient for cool stars with their deep convections layers and inefficient for hot stars that lack substantial convective envelopes \citep{2010ApJ...718L.145W,2010ApJ...719..602S}. 
For viscous dissipation in a convective layer, the timescale for spin-orbit alignment is proportional to both the sixth power of the scaled orbital separation, ($a$/\rstar)$^6$, and the square of the star-to-planet mass ratio, (\mstar/\mplanet)$^{2}$, \citep{1977A&A....57..383Z}. 
Thus, if short-orbit, giant planets migrate via high-eccentricty pathways and realign via tidal dissipation then, for cool stars, there should be a trend from spin-orbit alignment for close-in planets to a broad range of obliquities for those in more distant orbits. 

We selected those cool-star systems with $T_{\rm eff} \leq 6150$ K (to account for uncertainties) and with a mean uncertainty on $\lambda$ less than $20^\circ$. 
Their spin-orbit angles are plotted in Figure~\ref{fig:obliq-dist}, as a function of orbital distance in units of stellar radii $a/R_\star$, which is determined directly from the depth and width of the transit lightcurve \citep{2003ApJ...585.1038S}. We see that $\lambda$ is confined to within $\sim 20^\circ$ of aligned at orbital separations of $\lesssim$ 15 stellar radii and that the distribution is broad at greater separations. 
No such pattern is observed for hot stars, which exhibit a broad distribution of $\lambda$ at $a/R_\star < 15$. 
This suggests that a high fraction of hot Jupiters orbiting cool stars used to be misaligned and that tides changed that initial distribution either by realigning the orbits or by destroying misaligned planets.

\subsection{Giant-planet migration}
\begin{figure}
\includegraphics[width=90mm]{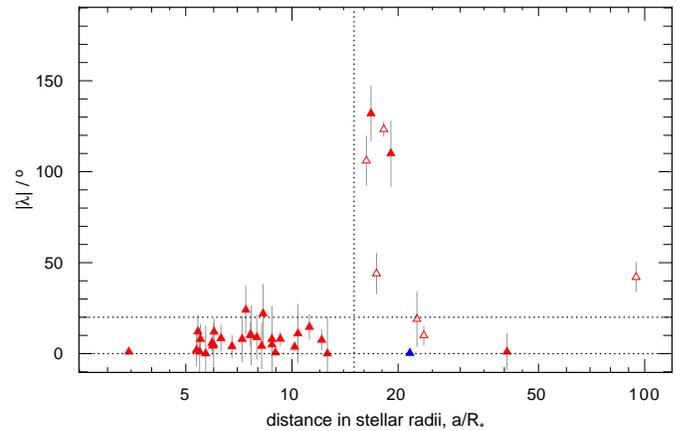}
\caption{The projected obliquity, $\lambda$, of planetary orbital planes as a function of scaled orbital distance, $a$/\rstar, for those systems with \teff\ $\leq$ 6150 K and $\overline{\sigma}_{\rm \lambda} < 20^\circ$. 
The filled symbols depict near-circular orbits ($e < 0.1$ or $e$ consistent with zero) and the open symbols depict eccentric orbits. WASP-84b is depicted by a blue triangle.
\label{fig:obliq-dist}} 
\end{figure}

We note that orbits shorter than $a/R_\star < 15$ are circular, whereas longer orbits are often eccentric (Figure~\ref{fig:obliq-dist}). 
As tides become ineffective with sufficient distance from the star, the population that we are beginning to see at $a/R_\star > 15$ may be close to the initial distribution in spin-orbit angle. 
The misaligned and/or eccentric systems are suggestive of high-eccentricity migration and the aligned, circular system of WASP-84 is suggestive of disc-driven migration. 
Thus both pathways appear to factor in the inward migration of giant planets; 
by measuring more systems in the weak-tide regime we can determine their relative contributions.

\section*{Acknowledgements}
The research leading to these results has received funding from the European Community's Seventh Framework Programme (FP7/2013-2016) under grant agreement number 312430 (OPTICON).
The Liverpool Telescope is operated on the island of La Palma by Liverpool John Moores University in the Spanish Observatorio del Roque de los Muchachos of the Instituto de Astrofisica de Canarias with financial support from the UK Science and Technology Facilities Council. 
A. H.\,M.\,J. Triaud is a Swiss National Science Foundation fellow under grant number P300P2-147773. 
John Southworth maintains a catalogue of physical properties, including the obliquities, of transiting planetary systems at \url{http://www.astro.keele.ac.uk/jkt/tepcat/}.

\end{document}